\def\aeti{$\alpha$-(BEDT\--TTF)$_2$I$_3$}
\def\cm{cm$^{-1}$}
\documentclass[aps,prl,lengthcheck,superscriptaddress,showpacs]{revtex4-1}
\usepackage{graphicx}%
\usepackage{color}
\usepackage{epstopdf}

\begin{document}
\title{Electronic Correlations Among the Dirac Electrons in $\alpha$-(BEDT-TTF)$_2$I$_3$\\
Unveiled by High-Pressure Optical Spectroscopy}

\author{Weiwu Li}
\affiliation{1.~Physikalisches Institut, Universit\"at Stuttgart, D-70569 Stuttgart, Germany}
\author{Ece Uykur}
\email{ece.uykur@pi1.physik.uni-stuttgart.de}
\affiliation{1.~Physikalisches Institut, Universit\"at Stuttgart, D-70569 Stuttgart, Germany} \affiliation{Experimentalphysik 2, Universit\"{a}t Augsburg, D-86159 Augsburg, Germany}
\author{Christine A. Kuntscher}
\affiliation{Experimentalphysik 2, Universit\"{a}t Augsburg, D-86159 Augsburg, Germany}
\author{Martin Dressel}
\affiliation{1.~Physikalisches Institut, Universit\"at Stuttgart, D-70569 Stuttgart, Germany}

\begin{abstract}
The charge-ordered insulator $\alpha$-(BEDT-TTF)$_2$I$_3$ gradually evolves to a metal
when pressure is applied,
and at low temperatures the electronic bands form tilted Dirac-like cones.
A metallic state with a frequency-independent optical conductivity indicates
the coexistence of the trivial and massless Dirac electrons in the system.
Our infrared investigations also reveal  that at the boundary between insulating and metallic states an energy gap opens due to correlated massive Dirac fermions, which is gradually suppressed when pressure increases.
\end{abstract}

\maketitle

Among the class of quasi two-dimensional organic conductors, \aeti\ is probably the most studied compound thanks to its rich phase diagram. At ambient conditions a pronounced
and sharp metal-insulator transition is observed at $T_{\rm CO}=135$~K \cite{Bender84b}
that eventually was explained by electronic charge order \cite{Kino95}.
With external pressure the insulator can be tuned to a zero-gap electronic state,
and even superconductivity is reached \cite{Kajita14}.
\aeti\ is the first realization of a two-dimensional multilayer massless Dirac fermion
bulk system \cite{Katayama06}; in many aspects its properties are distinct from two-dimensional graphene.
Apart from extensive studies on the ambient-pressure charge-ordered state \cite{Tomic15},
recently the focus shifted to the exotic properties of the Dirac electrons
that occur under high pressure.
The Dirac-fermion state has been proposed via different theoretical approaches \cite{Katayama06,Kino06,Kobayashi07,Alemany12,Suzumura12} and confirmed by various experimental methods such as dc transport \cite{Tajima07,Liu16}, interlayer magnetoresistance \cite{Tajima09,Tajima10}, quantum Hall effect \cite{Tajima12}, optical \cite{Beyer16,Peterseim16} and NMR spectroscopy \cite{Hirata16,Hirata17}.

Dirac-electron systems are subject to strong electronic correlations due to the unscreened long-range Coulomb interaction arising with the vanishing density of state around the Dirac point \cite{Kotov12}. The correlation strength can be controlled by the Coulomb coupling constant $\alpha\approx e^{2}/\epsilon\hbar v_F$, which is the ratio of the Coulomb potential and the electron kinetic energy ($\epsilon$ is the dielectric constant and $v_F$ is the Fermi velocity). It was predicted that when $\alpha$ is above some critical value, the massless Dirac electrons become massive with a gap opening near the Dirac point \cite{Knveshchenko09}; a gap size of 100~meV was indeed observed, for instance, in graphene \cite{Zhang08}.
In $\alpha$-(BEDT-TTF)$_2$I$_3$, $v_F$ is one order of magnitude smaller
compared to graphene \cite{Monteverde13}, therefore stronger interactions are expected,
which have been experimentally shown as an anomalous upturn in dc resistance $\rho(T)$
\cite{Liu16} and spin susceptibility \cite{Hirata17}, and strongly modified $v_F$ \cite{Hirata16}.

Optical spectroscopy is well suited to study gaps in the electronic structure and to learn about
the responsible interactions. The appearance of Dirac nodes in optical spectra has been discussed for two- and three-dimensional systems theoretically and shown experimentally on several examples \cite{Mak08,Kuzmenko08}.
The situation is expected to be more complex in \aeti\ due to the presence
of tilted Dirac cones \cite{Kajita14}; nevertheless optical investigations
should give valuable insight into the nature of the Dirac state and the role of the electron-electron interactions.

To that end we have conducted high-pressure infrared studies on \aeti\ single crystals. The $T$-dependent reflectivity from 100 to 8000~\cm\ was measured in a type-IIa diamond anvil cell \cite{Keller77} utilizing a home-built setup \cite{Kuntscher14} operating
from room temperature down to 6~K. The pressure in the cell was determined for each $T$ {\it in situ} by the ruby luminescence method \cite{Mao86}. The optical conductivity $\sigma_1(\omega)$ is obtained via Kramers-Kronig analysis \cite{Pashkin06}. Further details regarding the samples and measurements are described in the Supplemental Material \cite{SM}.

\begin{figure}[h]%
\centering
\includegraphics[scale=1]{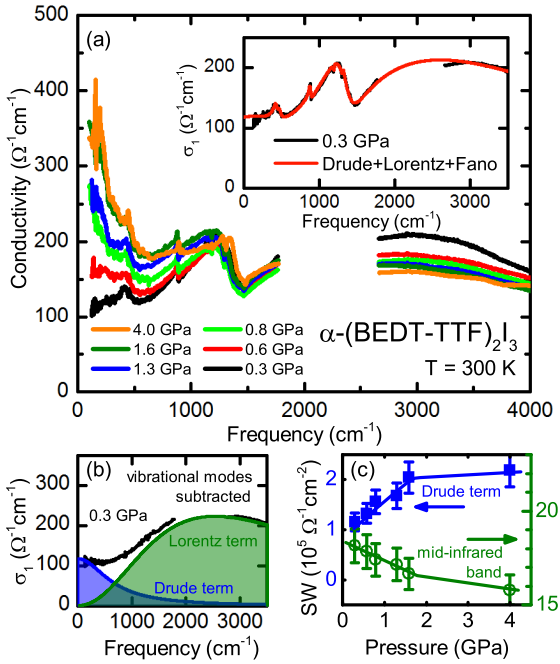}%
\caption{(a)~Pressure-evolution of the room-temperature optical conductivity of \aeti.
The inset displays the Drude-Lorentz-Fano fit to the 0.3~GPa spectrum as an example.
(b)~After subtracting the vibrational modes,
$\sigma_1(\omega)$ can be decomposed into a Drude term and mid-infrared contributions described by a Lorentz term.
(c)~Pressure dependence of the spectral weight (SW) for both components at $T=300$~K; the blue squares (Drude) refers to the left axis, the open green dots (Lorentz) correspond to the right axis.
\label{RT_OC}
}
\end{figure}
Fig.~\ref{RT_OC} displays the room-temperature optical conductivity of \aeti\ for various pressures. The overall shape and absolute value of $\sigma_1(\omega)$ measured at lowest pressure
is very similar to previous results recorded under ambient conditions \cite{Dressel94,Beyer16}, indicating the high quality and reproducibility of our experiments.
The spectrum at $p=0.3$~GPa consists of a strong mid-infrared absorption band around 3000~cm$^{-1}$ and an overdamped Drude contribution.
In addition to the electronic excitations,
several vibrational features are present between 400 and 1300~\cm, originating from the electron-molecular vibrational-coupled A$_g$ modes \cite{Dressel04}.
Since these modes obscure the electronic background,
we fitted our conductivity spectra by Drude, Lorentz, and Fano terms; an example is given in the inset of Fig.~\ref{RT_OC}(a).
For the further analysis these vibrations have been subtracted as plotted in Fig.~\ref{RT_OC}(b).
Below 1000~\cm\ $\sigma_1(\omega)$ becomes enhanced by pressure
with a progressive change from a broad to a narrower Drude behavior above 0.8~GPa,
suggesting a bad- to good-metal transition, i.e.\ from incoherent to more coherent transport.
The mid-infrared band persists up to pressure as high as 4.0~GPa
with a gradual suppression in intensity.
The enhancement of the low-frequency $\sigma_1(\omega)$ is in line with our $p$-dependent dc-resistivity measurements \cite{Beyer16}, suggesting that even at room temperature the electronic state changes with applied pressure.

To characterize the charge dynamics under pressure quantitatively,
the spectral weight of the zero-frequency and mid-infrared bands
is plotted in Fig.~\ref{RT_OC}(c).
With increasing pressure a systematic transfer of spectral weight occurs from high energies to the Drude component; this trend seems to saturate above $p\approx 2$~GPa.
The energy range that the spectral weight transfer is observed over a large energy range is comparable to the on-site Coulomb repulsion $U\approx~0.4$~eV \cite{Kobayashi07}; this is reminiscent of the behavior observed in the bandwidth-controlled metal-insulator in Mott insulators \cite{Basov11} and also occurs in other charge-ordered organic compounds \cite{Dressel04,Drichko06}. The mid-infrared conductivity arising from transitions between
the bands split by electronic correlations; the enhancement of the Drude-like response corresponds to the overlap of the two bands that increases with  pressure \cite{Rozenberg95}. Hence our room-temperature optical studies provide strong evidence that electronic correlations are important for \aeti, in accord with recent dc transport \cite{Liu16} and NMR measurements \cite{Hirata16, Hirata17}; our findings are consistent with conclusions based on the extended Hubbard model, 
which considers on-site and  inter-site Coulomb repulsion \cite{Kobayashi07,Kobayashi09,Suzumura12}.

\begin{figure}%
\centering
\includegraphics[scale=1]{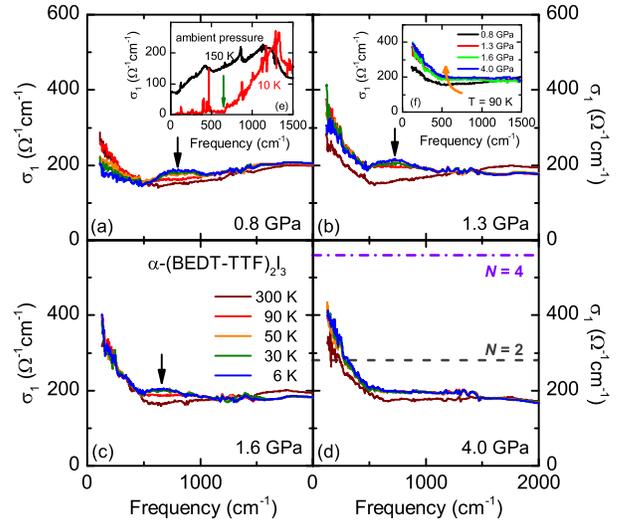}%
\caption{(a-d) Temperature-dependent conductivity at $p = 0.8´$, 1.3, 1.6, and 4.0~GPa.
The black arrows mark the absorption feature due to excitations across an energy gap.
In panel (d) the bulk conductance corresponding to $G(\omega)/c$ for $N=2$ and $4$
is indicated by dashed gray and dash-dotted purple lines, respectively,
where $c\approx$1.7~nm is the $c$-axis lattice constant \cite{Bender84b,Kondo09}.
(e)~The ambient-pressure $\sigma_1(\omega)$ above and below the metal-insulator transition $T_{\rm CO}$ clearly shows the opening of the charge-order gap (green arrow).
(f)~$\sigma_1(\omega, T=90~{\rm K})$ for various pressures;
between $p=0.8$ and 1.3~GPa $\sigma_1(\omega)$ becomes frequency independent above 500~\cm.}%
\label{tempdep}%
\end{figure}

The $T$-dependent conductivity spectra are displayed in Fig.~\ref{tempdep} at various pressures. Previously we showed \cite{Beyer16} that the low-pressure regime (0-0.8~GPa at 8~K) is characterized as a charge-ordered ground state at low temperatures: The non-zero but relatively small spectral weight below 400~\cm\ at $T=150$~K is suppressed when the metal-insulator transition is reached at $T_{\rm CO}$ with a clear gap opening, marked by the green arrow in Fig.~\ref{tempdep}(e). On the other hand, the temperature evolutions of $\sigma_1(\omega)$ at pressures above 0.8~GPa are qualitatively different. Taking the conductivity at $0.8$~GPa [Fig.~\ref{tempdep}(a)], for instance, \aeti\ exhibits metallic behavior with a characteristic spectral weight transfer
towards low energies as $T$  decreases, and a Drude-like response below 500~\cm.
The energy range of this transfer ($\sim$0.5 eV) suggests  strong modifications of the trivial bands (non-Dirac bands, possible possess parabolic dispersion) upon pressure. This metallic behavior persists up to the maximum pressure measured (4.0~GPa), indicating that the ground state above $p \approx 0.8$~GPa is distinct from the charge-ordered state; obviously we have entered a region where Dirac electrons exist;
in accord with previous dc transport measurements \cite{Liu16}.

A close inspection of $\sigma_1(\omega,T)$ at $p=0.8$~GPa
reveals a gradual transition at low temperatures.
Since the vibrational features at around 1300~\cm\ exhibit a rather large intensity,
we allocate the pressure of 0.8~GPa close to the boundary between the insulator and Dirac metal state \cite{SM}.
As shown in Fig.~\ref{tempdep}(a), the low-frequency conductivity is suppressed below 90~K,
and a pronounced peak develops at around 800~\cm\ that we assign to excitations across an energy gap and defined as a ``pseudogap" openning. With increasing pressure, this structure shifts to lower energies and becomes gradually suppressed. We can unambiguously trace the gap feature up to $p=1.6$~GPa; but at 4.0~GPa we cannot resolve the band anymore within the accessible temperature and energy range. Our discovery of a metallic ground state with a gap feature is in line with the anomalous upturn observed by pressure-dependent dc measurements \cite{Liu16}.
A similar behavior is observed in density-wave systems with an only partially-gapped Fermi surface \cite{Perucchi04,Barisic10}, or Mott insulators with metallic carriers remaining
\cite{Dumm09,Pustogow17}; hence we have to discuss the origin of the peak structure and in-gap absorption in more detail.

Two-dimensional massless Dirac electrons manifest themselves in a frequency-independent conductivity; the interband conductivity per layer should be a universal constant
$G(\omega) = N\pi G_0/4$, with $N$ the number of non-degenerate Dirac cones and $G_0=2e^2/\hbar$ is the quantum conductance. Such peculiar $\sigma_1(\omega)$ behavior has been discussed theoretically and experimentally, for instance, in monolayer graphene \cite{Mak08} and quasi-two-dimensional graphite \cite{Kuzmenko08}.
As displayed in Fig.~\ref{tempdep}(f) $\sigma_1(\omega)$ becomes constant
between 400 and 1200~\cm\ as pressure rises from 0.8 to 1.3~GPa and then remains unchanged.
The flat conductivity observed in this large spectral range is taken as strong evidence
for massless Dirac fermions;
the observed value is quite close to the predicted conductance for $N=2$ [Fig.~\ref{tempdep}(d)]. In comparison to the case of graphite and graphene, the energy region of the
$\omega$-independent conductivity is limited due to the bandstructure,
in agreement with theory \cite{Kobayashi09} and NMR experiments \cite{Hirata16,Hirata17}.

With the modifications of the trivial bands and the existence of the Dirac state
it is rather challenging to analyze the low-energy behavior of $\sigma_1(\omega)$.
Since the Drude component below 300~\cm\ does not show a significant $T$-dependence
above 1.3~GPa, we suggest two contributions:
(i)~the correlated massive electrons and (ii)~the thermally excited massless Dirac electrons. The former contribution should become stronger as the temperature decreases.
For a pure Dirac system with the chemical potential located exactly at the Dirac point,
the spectral weight of the zero-energy response should decrease upon cooling \cite{Kuzmenko08}.
Magnetotransport and optical measurements \cite{Monteverde13,Beyer16,Peterseim16}
indicate the coexistence of massive and Dirac carriers. This coexistence can also explain the deviations from the quantum resistence observed in $\rho(T)$ between 300 and 10~K \cite{Tajima07,Liu16,Beyer16}.
The tilting of the Dirac cones present in \aeti\ \cite{Hirata16,Hirata17}
leads to a strong anisotropy of the Fermi velocity and modifies the interband transitions of the Dirac electrons. In accord with Suzumura {\it et al.} \cite{Suzumura14} we expect an effect only in the THz optical response (around 20~\cm), and not at higher energies where we observe $\sigma_1(\omega) = {\rm const.}$

\begin{figure}%
\centering
\includegraphics[scale=1]{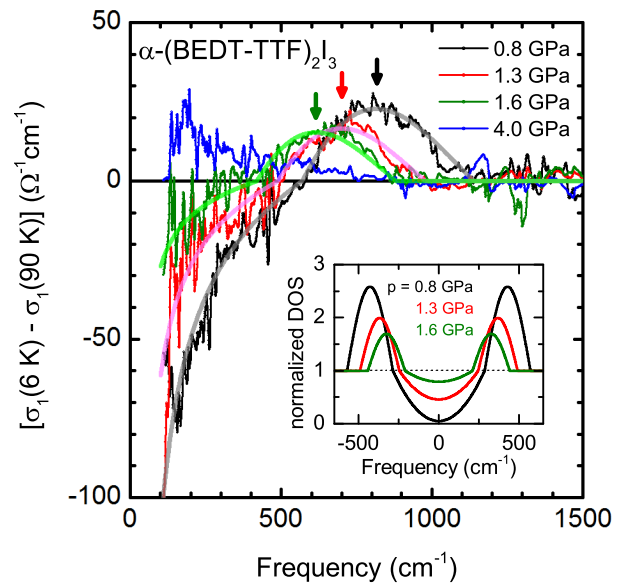}%
\caption{Pressure dependence of the changes in the optical conductivity between $T=90$ and 6~K.
The peak structure indicated by the arrows shifts toward lower energies and diminishes as pressure increases. The solid lines correspond to $\Delta\sigma_1(\omega)$ calculated from the electronic density of states  given in the inset. The pseudogap closes by pressure and is completely absent at $p=4$~GPa.}%
\label{difference}%
\end{figure}

More recently, it was suggested that the massless Dirac electrons are strongly correlated due to the unscreened long-range Coulomb repulsion giving rise to an anomalous increase in $\rho(T)$ \cite{Liu16}, strong modification of the Fermi velocity \cite{Hirata16} and excitonic mass generation at low $T$ \cite{Hirata17}.
Our optical results not only provide strong evidence of interacting Dirac fermions,
but give insight into the actual density of electronic states.
For a better illustration, in Fig.~\ref{difference} we plot the difference of the conductivity between $T=6$ and 90~K.
The openning of a pseudogap is well identified as peaks in $\Delta\sigma_1(\omega)$ and
indicated by arrows.  From a proposed density of states as displayed in the inset for the different pressure values, a fitting to the difference of the optical conductivity (solid lines) can be obtained. In our model we assume the optical joint density of states to be proportional to $\epsilon_2=4\pi\sigma_1(\omega)/\omega$ and neglect the effect of the matrix element \cite{DresselGruner02} . At low temperatures ($T=6$~K) a pronounced pseudogap is present with a significant reduction of the density of states. The spectral weight missing below the gap is recovered at higher energies, which manifest itself with a spectral weight transfer from low to high energies resulting an absorption peak in the optical spectra. We defined the energy of this peak structure as $2\Delta$ (shown with arrows in Fig.~\ref{tempdep} and ~\ref{difference}), and for the pseudogap we obtain $\Delta$ = 410, 350, and 310~\cm\ at $p=0.8$, 1.3, and 1.6~GPa respectively; and it seemed to be completely suppressed at 4.0~GPa. As can be seen from the fittings in Fig.~\ref{difference},  with rising pressure, an assumption of the decrease of $\Delta$ and suppression of the pseudogap by gradual filling in states from above can reproduce the experimental finding satisfactorily.  

These findings can be compared to  the gap values estimated by Khveshchenko \cite{Knveshchenko09} using:
\begin{equation}
\Delta=v_F\hbar\Lambda\text{exp}[(-2\pi+ 4\arctan\sqrt{2\tilde{\alpha}-1})/(\sqrt{2\tilde{\alpha}-1})]/k_B \quad ,
\end{equation}
where $\tilde{\alpha}=\alpha/(1+N\pi\alpha/8\sqrt{2})$, 
and $\Lambda = 0.667~{\rm \AA}^{-1}$ is a momentum cutoff at the inverse lattice constant \cite{Kondo09}.
For our estimation the dielectric constant $\epsilon_{\infty}\approx4$ was directly extracted from the high frequency limit (supplementary materials); the Fermi velocity $v_F \approx 2.4 - 10 \times10^{4}~{\rm ms}^{-1}$ was taken from Refs.~\cite{Tajima07,Alemany12,Hirata16}. With the assumption of $N=2$, we obtain $\Delta\approx300$~\cm, in good agreement with our optical data.
There are two non-degerate Dirac cones ($N=4$) predicted \cite{Kino06,Alemany12},
which merge into one at high pressure ($N=2$) \cite{Kobayashi07}.
In combination of the observed universal constant conductance and the pseudogap, our present optical study yields strong evidence of a single Dirac cone.
In a theoretical study of uniaxial pressure \cite{Kishigi17} it was recently predicted that \aeti\ exhibits three-quarter Dirac points; further experiments have to show whether this is
of relevance for the optical conductivity.

\begin{figure}%
\centering
\includegraphics[scale=0.5]{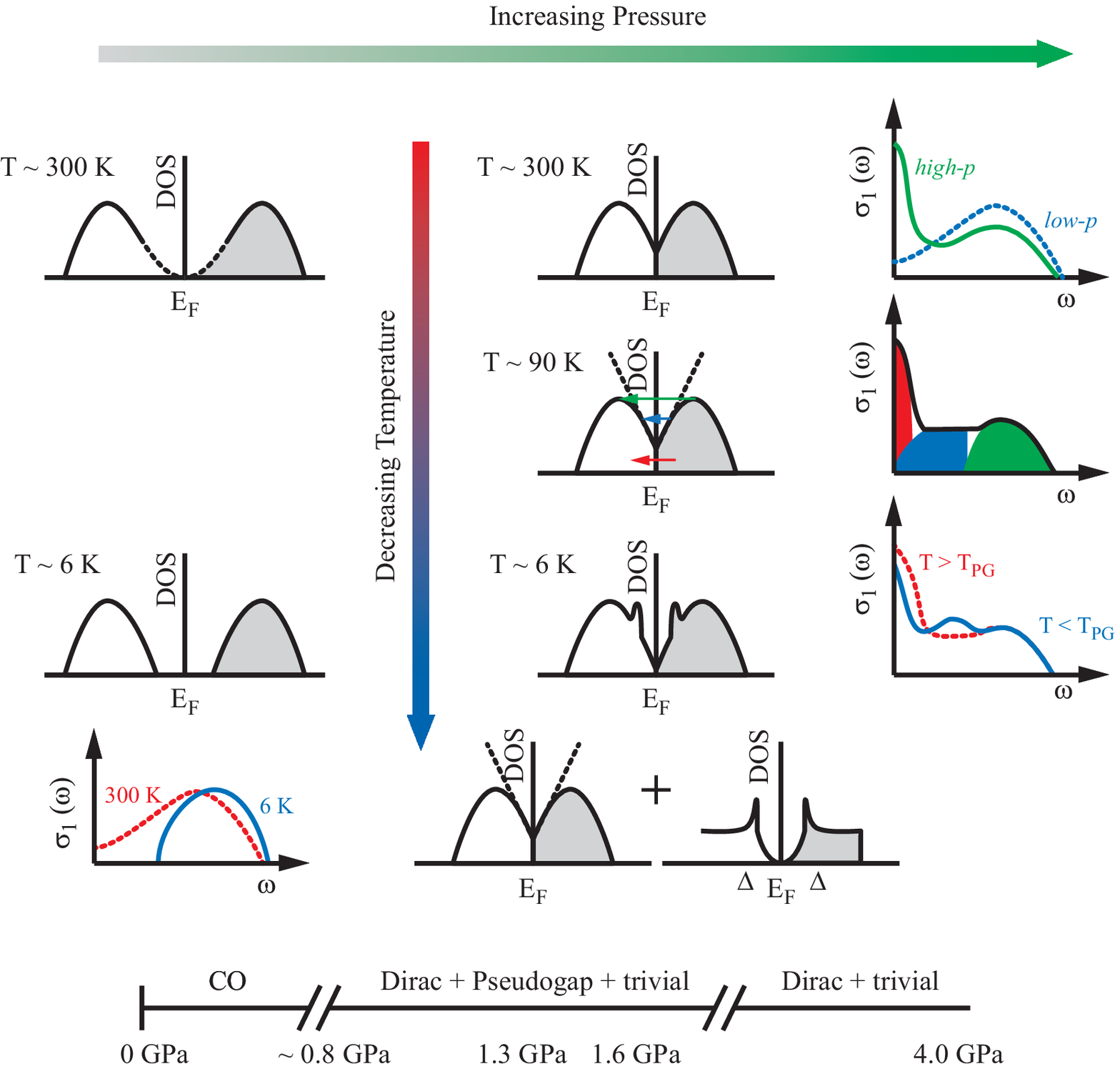}%
\caption{ The schematic diagrams for the electronic structure and corresponding optical conductivity of \aeti\ for the low-pressure insulator state and the high-pressure Dirac state
at various temperatures. Here T$_{PG}$ stands for the temperature, where the pseudogap starts to open. At the bottom the pressure evolution of the various electronic phases of \aeti\ at low temperatures is summarized. For a detailed description see Conclusions. }%
\label{schematic}%
\end{figure}

The conclusions drawn from our pressure-dependent optical investigations are summarized in the schematic density of states and $\sigma_1(\omega)$ sketched in Fig.~\ref{schematic}.
At room temperature and for low pressure, \aeti\ is a bad metal
with a very broad Drude contribution to $\sigma_1(\omega)$ originating from the thermally excited electrons at the Fermi energy. Upon cooling, these excitations freeze out and
a clear charge-order gap develops in the optical conductivity. With rising pressure
these bands get closer to each other and overlap, leading to a more-or-less narrow Drude contribution (dark green curve).
As we cool down, the edges of the two bands develop linear dispersions.
In the high-pressure range $\sigma_1(\omega)$ consists of three components:
(i) a low-energy Drude response (red area),
(ii) frequency-independent conductivity due to the Dirac electrons (blue area) and
(iii) mid-infrared band arising with the incoherent transitions due to on-site and inter-site Coulomb repulsion (green area).
With further cooling, electronic correlations cave in a pseudogap
with states piling up at the edges. As a result, the Drude spectral weight is transferred to finite energies, leading to local peaks in $\sigma_1(\omega)$ around 800~\cm\ (blue curve).
Note, the behavior is distinct from superconductivity, where the missing spectral weight condenses in a $\delta(\omega=0)$-peak \cite{DresselGruner02}.
The phase diagram as a function of external pressure displayed in the lower part of Fig.~\ref{schematic} can be divided into three parts:
(i)~the charge order insulating state at low pressure;
(ii)~metallic states in the intermediate pressure regime consisting of massless Dirac electrons, next to carriers in correlation-split and trivial bands;
(iii)~above 4.0~GPa only the Dirac electronic state and carriers in trivial bands remain (dashed red curve).
Our optical results demonstrate clear fingerprints of the electronic correlations between the Dirac electrons; the interaction can be tuned by temperature and pressure.
We call for complementary spectroscopic experiments to directly confirm our findings and
further efforts for a theoretical description of our observations.

\begin{acknowledgments}
We thank D. Schweitzer for providing the single crystals, G. Untereiner for technical support and M. Hirata for discussions. We also acknowledge the support by the Deutsche Forschungsgemeinschaft (DFG). E.U acknowledge the supported by ``Margarete von Wrangell Habilitation Program" by the Ministry of Sciences, Research, and Arts in Baden Württemberg. E. U. and C. K. acknowledge financial support by the Federal Ministry of Education and Research (BMBF), Germany, through Grant No. 05K13WA1 (Verbundprojekt 05K2013, Teilprojekt 1, PT-DESY).
\end{acknowledgments}

\bibliographystyle{apsrev4-1}
\bibliography{References}

\begin{thebibliography}{43}%
\makeatletter
\providecommand \@ifxundefined [1]{%
 \@ifx{#1\undefined}
}%
\providecommand \@ifnum [1]{%
 \ifnum #1\expandafter \@firstoftwo
 \else \expandafter \@secondoftwo
 \fi
}%
\providecommand \@ifx [1]{%
 \ifx #1\expandafter \@firstoftwo
 \else \expandafter \@secondoftwo
 \fi
}%
\providecommand \natexlab [1]{#1}%
\providecommand \enquote  [1]{``#1''}%
\providecommand \bibnamefont  [1]{#1}%
\providecommand \bibfnamefont [1]{#1}%
\providecommand \citenamefont [1]{#1}%
\providecommand \href@noop [0]{\@secondoftwo}%
\providecommand \href [0]{\begingroup \@sanitize@url \@href}%
\providecommand \@href[1]{\@@startlink{#1}\@@href}%
\providecommand \@@href[1]{\endgroup#1\@@endlink}%
\providecommand \@sanitize@url [0]{\catcode `\\12\catcode `\$12\catcode
  `\&12\catcode `\#12\catcode `\^12\catcode `\_12\catcode `\%12\relax}%
\providecommand \@@startlink[1]{}%
\providecommand \@@endlink[0]{}%
\providecommand \url  [0]{\begingroup\@sanitize@url \@url }%
\providecommand \@url [1]{\endgroup\@href {#1}{\urlprefix }}%
\providecommand \urlprefix  [0]{URL }%
\providecommand \Eprint [0]{\href }%
\providecommand \doibase [0]{http://dx.doi.org/}%
\providecommand \selectlanguage [0]{\@gobble}%
\providecommand \bibinfo  [0]{\@secondoftwo}%
\providecommand \bibfield  [0]{\@secondoftwo}%
\providecommand \translation [1]{[#1]}%
\providecommand \BibitemOpen [0]{}%
\providecommand \bibitemStop [0]{}%
\providecommand \bibitemNoStop [0]{.\EOS\space}%
\providecommand \EOS [0]{\spacefactor3000\relax}%
\providecommand \BibitemShut  [1]{\csname bibitem#1\endcsname}%
\let\auto@bib@innerbib\@empty
\bibitem [{\citenamefont {Bender}\ \emph {et~al.}(1984)\citenamefont {Bender},
  \citenamefont {Dietz}, \citenamefont {Endres}, \citenamefont {Helberg},
  \citenamefont {Hennig}, \citenamefont {Keller}, \citenamefont {Schäfer},\
  and\ \citenamefont {Schweitzer}}]{Bender84b}%
  \BibitemOpen
  \bibfield  {author} {\bibinfo {author} {\bibfnamefont {K.}~\bibnamefont
  {Bender}}, \bibinfo {author} {\bibfnamefont {K.}~\bibnamefont {Dietz}},
  \bibinfo {author} {\bibfnamefont {H.}~\bibnamefont {Endres}}, \bibinfo
  {author} {\bibfnamefont {H.~W.}\ \bibnamefont {Helberg}}, \bibinfo {author}
  {\bibfnamefont {I.}~\bibnamefont {Hennig}}, \bibinfo {author} {\bibfnamefont
  {H.~J.}\ \bibnamefont {Keller}}, \bibinfo {author} {\bibfnamefont {H.~W.}\
  \bibnamefont {Schäfer}}, \ and\ \bibinfo {author} {\bibfnamefont
  {D.}~\bibnamefont {Schweitzer}},\ }\href {\doibase 10.1080/00268948408072071}
  {\bibfield  {journal} {\bibinfo  {journal} {Molecular Crystals and Liquid
  Crystals}\ }\textbf {\bibinfo {volume} {107}},\ \bibinfo {pages} {45}
  (\bibinfo {year} {1984})},\ \Eprint
  {http://arxiv.org/abs/https://doi.org/10.1080/00268948408072071}
  {https://doi.org/10.1080/00268948408072071} \BibitemShut {NoStop}%
\bibitem [{\citenamefont {Kino}\ and\ \citenamefont {Fukuyama}(1995)}]{Kino95}%
  \BibitemOpen
  \bibfield  {author} {\bibinfo {author} {\bibfnamefont {H.}~\bibnamefont
  {Kino}}\ and\ \bibinfo {author} {\bibfnamefont {H.}~\bibnamefont
  {Fukuyama}},\ }\href {\doibase 10.1143/JPSJ.64.1877} {\bibfield  {journal}
  {\bibinfo  {journal} {Journal of the Physical Society of Japan}\ }\textbf
  {\bibinfo {volume} {64}},\ \bibinfo {pages} {1877} (\bibinfo {year}
  {1995})},\ \Eprint
  {http://arxiv.org/abs/https://doi.org/10.1143/JPSJ.64.1877}
  {https://doi.org/10.1143/JPSJ.64.1877} \BibitemShut {NoStop}%
\bibitem [{\citenamefont {Kajita}\ \emph {et~al.}(2014)\citenamefont {Kajita},
  \citenamefont {Nishio}, \citenamefont {Tajima}, \citenamefont {Suzumura},\
  and\ \citenamefont {Kobayashi}}]{Kajita14}%
  \BibitemOpen
  \bibfield  {author} {\bibinfo {author} {\bibfnamefont {K.}~\bibnamefont
  {Kajita}}, \bibinfo {author} {\bibfnamefont {Y.}~\bibnamefont {Nishio}},
  \bibinfo {author} {\bibfnamefont {N.}~\bibnamefont {Tajima}}, \bibinfo
  {author} {\bibfnamefont {Y.}~\bibnamefont {Suzumura}}, \ and\ \bibinfo
  {author} {\bibfnamefont {A.}~\bibnamefont {Kobayashi}},\ }\href {\doibase
  10.7566/JPSJ.83.072002} {\bibfield  {journal} {\bibinfo  {journal} {Journal
  of the Physical Society of Japan}\ }\textbf {\bibinfo {volume} {83}},\
  \bibinfo {pages} {072002} (\bibinfo {year} {2014})},\ \Eprint
  {http://arxiv.org/abs/http://dx.doi.org/10.7566/JPSJ.83.072002}
  {http://dx.doi.org/10.7566/JPSJ.83.072002} \BibitemShut {NoStop}%
\bibitem [{\citenamefont {Katayama}\ \emph {et~al.}(2006)\citenamefont
  {Katayama}, \citenamefont {Kobayashi},\ and\ \citenamefont
  {Suzumura}}]{Katayama06}%
  \BibitemOpen
  \bibfield  {author} {\bibinfo {author} {\bibfnamefont {S.}~\bibnamefont
  {Katayama}}, \bibinfo {author} {\bibfnamefont {A.}~\bibnamefont {Kobayashi}},
  \ and\ \bibinfo {author} {\bibfnamefont {Y.}~\bibnamefont {Suzumura}},\
  }\href {\doibase 10.1143/JPSJ.75.054705} {\bibfield  {journal} {\bibinfo
  {journal} {J. Phys. Soc. Jpn.}\ }\textbf {\bibinfo {volume} {75}},\ \bibinfo
  {pages} {054705} (\bibinfo {year} {2006})},\ \Eprint
  {http://arxiv.org/abs/http://dx.doi.org/10.1143/JPSJ.75.054705}
  {http://dx.doi.org/10.1143/JPSJ.75.054705} \BibitemShut {NoStop}%
\bibitem [{\citenamefont {Tomi{\'c}}\ and\ \citenamefont
  {Dressel}(2015)}]{Tomic15}%
  \BibitemOpen
  \bibfield  {author} {\bibinfo {author} {\bibfnamefont {S.}~\bibnamefont
  {Tomi{\'c}}}\ and\ \bibinfo {author} {\bibfnamefont {M.}~\bibnamefont
  {Dressel}},\ }\href {http://stacks.iop.org/0034-4885/78/i=9/a=096501}
  {\bibfield  {journal} {\bibinfo  {journal} {Rep. Progr. Phys.}\ }\textbf
  {\bibinfo {volume} {78}},\ \bibinfo {pages} {096501} (\bibinfo {year}
  {2015})}\BibitemShut {NoStop}%
\bibitem [{\citenamefont {Kino}\ and\ \citenamefont {Miyazaki}(2006)}]{Kino06}%
  \BibitemOpen
  \bibfield  {author} {\bibinfo {author} {\bibfnamefont {H.}~\bibnamefont
  {Kino}}\ and\ \bibinfo {author} {\bibfnamefont {T.}~\bibnamefont
  {Miyazaki}},\ }\href {\doibase 10.1143/JPSJ.75.034704} {\bibfield  {journal}
  {\bibinfo  {journal} {J. Phys. Soc. Jpn.}\ }\textbf {\bibinfo {volume}
  {75}},\ \bibinfo {pages} {034704} (\bibinfo {year} {2006})},\ \Eprint
  {http://arxiv.org/abs/http://dx.doi.org/10.1143/JPSJ.75.034704}
  {http://dx.doi.org/10.1143/JPSJ.75.034704} \BibitemShut {NoStop}%
\bibitem [{\citenamefont {Kobayashi}\ \emph {et~al.}(2007)\citenamefont
  {Kobayashi}, \citenamefont {Katayama}, \citenamefont {Suzumura},\ and\
  \citenamefont {Fukuyama}}]{Kobayashi07}%
  \BibitemOpen
  \bibfield  {author} {\bibinfo {author} {\bibfnamefont {A.}~\bibnamefont
  {Kobayashi}}, \bibinfo {author} {\bibfnamefont {S.}~\bibnamefont {Katayama}},
  \bibinfo {author} {\bibfnamefont {Y.}~\bibnamefont {Suzumura}}, \ and\
  \bibinfo {author} {\bibfnamefont {H.}~\bibnamefont {Fukuyama}},\ }\href
  {\doibase 10.1143/JPSJ.76.034711} {\bibfield  {journal} {\bibinfo  {journal}
  {J. Phys. Soc. Jpn.}\ }\textbf {\bibinfo {volume} {76}},\ \bibinfo {pages}
  {034711} (\bibinfo {year} {2007})},\ \Eprint
  {http://arxiv.org/abs/http://dx.doi.org/10.1143/JPSJ.76.034711}
  {http://dx.doi.org/10.1143/JPSJ.76.034711} \BibitemShut {NoStop}%
\bibitem [{\citenamefont {Alemany}\ \emph {et~al.}(2012)\citenamefont
  {Alemany}, \citenamefont {Pouget},\ and\ \citenamefont
  {Canadell}}]{Alemany12}%
  \BibitemOpen
  \bibfield  {author} {\bibinfo {author} {\bibfnamefont {P.}~\bibnamefont
  {Alemany}}, \bibinfo {author} {\bibfnamefont {J.-P.}\ \bibnamefont {Pouget}},
  \ and\ \bibinfo {author} {\bibfnamefont {E.}~\bibnamefont {Canadell}},\
  }\href {\doibase 10.1103/PhysRevB.85.195118} {\bibfield  {journal} {\bibinfo
  {journal} {Phys. Rev. B}\ }\textbf {\bibinfo {volume} {85}},\ \bibinfo
  {pages} {195118} (\bibinfo {year} {2012})}\BibitemShut {NoStop}%
\bibitem [{\citenamefont {Suzumura}\ and\ \citenamefont
  {Kobayashi}(2012)}]{Suzumura12}%
  \BibitemOpen
  \bibfield  {author} {\bibinfo {author} {\bibfnamefont {Y.}~\bibnamefont
  {Suzumura}}\ and\ \bibinfo {author} {\bibfnamefont {A.}~\bibnamefont
  {Kobayashi}},\ }\href@noop {} {\bibfield  {journal} {\bibinfo  {journal}
  {Crystals}\ }\textbf {\bibinfo {volume} {2}},\ \bibinfo {pages} {266}
  (\bibinfo {year} {2012})}\BibitemShut {NoStop}%
\bibitem [{\citenamefont {Tajima}\ \emph {et~al.}(2007)\citenamefont {Tajima},
  \citenamefont {Sugawara}, \citenamefont {Tamura}, \citenamefont {Kato},
  \citenamefont {Nishio},\ and\ \citenamefont {Kajita}}]{Tajima07}%
  \BibitemOpen
  \bibfield  {author} {\bibinfo {author} {\bibfnamefont {N.}~\bibnamefont
  {Tajima}}, \bibinfo {author} {\bibfnamefont {S.}~\bibnamefont {Sugawara}},
  \bibinfo {author} {\bibfnamefont {M.}~\bibnamefont {Tamura}}, \bibinfo
  {author} {\bibfnamefont {R.}~\bibnamefont {Kato}}, \bibinfo {author}
  {\bibfnamefont {Y.}~\bibnamefont {Nishio}}, \ and\ \bibinfo {author}
  {\bibfnamefont {K.}~\bibnamefont {Kajita}},\ }\href {\doibase
  10.1209/0295-5075/80/47002} {\bibfield  {journal} {\bibinfo  {journal} {EPL}\
  }\textbf {\bibinfo {volume} {80}},\ \bibinfo {pages} {47002} (\bibinfo {year}
  {2007})}\BibitemShut {NoStop}%
\bibitem [{\citenamefont {Liu}\ \emph {et~al.}(2016)\citenamefont {Liu},
  \citenamefont {Ishikawa}, \citenamefont {Takehara}, \citenamefont {Miyagawa},
  \citenamefont {Tamura},\ and\ \citenamefont {Kanoda}}]{Liu16}%
  \BibitemOpen
  \bibfield  {author} {\bibinfo {author} {\bibfnamefont {D.}~\bibnamefont
  {Liu}}, \bibinfo {author} {\bibfnamefont {K.}~\bibnamefont {Ishikawa}},
  \bibinfo {author} {\bibfnamefont {R.}~\bibnamefont {Takehara}}, \bibinfo
  {author} {\bibfnamefont {K.}~\bibnamefont {Miyagawa}}, \bibinfo {author}
  {\bibfnamefont {M.}~\bibnamefont {Tamura}}, \ and\ \bibinfo {author}
  {\bibfnamefont {K.}~\bibnamefont {Kanoda}},\ }\href {\doibase
  10.1103/PhysRevLett.116.226401} {\bibfield  {journal} {\bibinfo  {journal}
  {Phys. Rev. Lett.}\ }\textbf {\bibinfo {volume} {116}},\ \bibinfo {pages}
  {226401} (\bibinfo {year} {2016})}\BibitemShut {NoStop}%
\bibitem [{\citenamefont {Tajima}\ and\ \citenamefont
  {Kajita}(2009)}]{Tajima09}%
  \BibitemOpen
  \bibfield  {author} {\bibinfo {author} {\bibfnamefont {N.}~\bibnamefont
  {Tajima}}\ and\ \bibinfo {author} {\bibfnamefont {K.}~\bibnamefont
  {Kajita}},\ }\href {\doibase 10.1088/1468-6996/10/2/024308} {\bibfield
  {journal} {\bibinfo  {journal} {Science and Technology of Advanced
  Materials}\ }\textbf {\bibinfo {volume} {10}},\ \bibinfo {pages} {024308}
  (\bibinfo {year} {2009})},\ \Eprint
  {http://arxiv.org/abs/http://dx.doi.org/10.1088/1468-6996/10/2/024308}
  {http://dx.doi.org/10.1088/1468-6996/10/2/024308} \BibitemShut {NoStop}%
\bibitem [{\citenamefont {Tajima}\ \emph {et~al.}(2010)\citenamefont {Tajima},
  \citenamefont {Sato}, \citenamefont {Sugawara}, \citenamefont {Kato},
  \citenamefont {Nishio},\ and\ \citenamefont {Kajita}}]{Tajima10}%
  \BibitemOpen
  \bibfield  {author} {\bibinfo {author} {\bibfnamefont {N.}~\bibnamefont
  {Tajima}}, \bibinfo {author} {\bibfnamefont {M.}~\bibnamefont {Sato}},
  \bibinfo {author} {\bibfnamefont {S.}~\bibnamefont {Sugawara}}, \bibinfo
  {author} {\bibfnamefont {R.}~\bibnamefont {Kato}}, \bibinfo {author}
  {\bibfnamefont {Y.}~\bibnamefont {Nishio}}, \ and\ \bibinfo {author}
  {\bibfnamefont {K.}~\bibnamefont {Kajita}},\ }\href {\doibase
  10.1103/PhysRevB.82.121420} {\bibfield  {journal} {\bibinfo  {journal} {Phys.
  Rev. B}\ }\textbf {\bibinfo {volume} {82}},\ \bibinfo {pages} {121420}
  (\bibinfo {year} {2010})}\BibitemShut {NoStop}%
\bibitem [{\citenamefont {Tajima}\ \emph {et~al.}(2012)\citenamefont {Tajima},
  \citenamefont {Kato}, \citenamefont {Sugawara}, \citenamefont {Nishio},\ and\
  \citenamefont {Kajita}}]{Tajima12}%
  \BibitemOpen
  \bibfield  {author} {\bibinfo {author} {\bibfnamefont {N.}~\bibnamefont
  {Tajima}}, \bibinfo {author} {\bibfnamefont {R.}~\bibnamefont {Kato}},
  \bibinfo {author} {\bibfnamefont {S.}~\bibnamefont {Sugawara}}, \bibinfo
  {author} {\bibfnamefont {Y.}~\bibnamefont {Nishio}}, \ and\ \bibinfo {author}
  {\bibfnamefont {K.}~\bibnamefont {Kajita}},\ }\href {\doibase
  10.1103/PhysRevB.85.033401} {\bibfield  {journal} {\bibinfo  {journal} {Phys.
  Rev. B}\ }\textbf {\bibinfo {volume} {85}},\ \bibinfo {pages} {033401}
  (\bibinfo {year} {2012})}\BibitemShut {NoStop}%
\bibitem [{\citenamefont {Beyer}\ \emph {et~al.}(2016)\citenamefont {Beyer},
  \citenamefont {Dengl}, \citenamefont {Peterseim}, \citenamefont {Wackerow},
  \citenamefont {Ivek}, \citenamefont {Pronin}, \citenamefont {Schweitzer},\
  and\ \citenamefont {Dressel}}]{Beyer16}%
  \BibitemOpen
  \bibfield  {author} {\bibinfo {author} {\bibfnamefont {R.}~\bibnamefont
  {Beyer}}, \bibinfo {author} {\bibfnamefont {A.}~\bibnamefont {Dengl}},
  \bibinfo {author} {\bibfnamefont {T.}~\bibnamefont {Peterseim}}, \bibinfo
  {author} {\bibfnamefont {S.}~\bibnamefont {Wackerow}}, \bibinfo {author}
  {\bibfnamefont {T.}~\bibnamefont {Ivek}}, \bibinfo {author} {\bibfnamefont
  {A.~V.}\ \bibnamefont {Pronin}}, \bibinfo {author} {\bibfnamefont
  {D.}~\bibnamefont {Schweitzer}}, \ and\ \bibinfo {author} {\bibfnamefont
  {M.}~\bibnamefont {Dressel}},\ }\href {\doibase 10.1103/PhysRevB.93.195116}
  {\bibfield  {journal} {\bibinfo  {journal} {Phys. Rev. B}\ }\textbf {\bibinfo
  {volume} {93}},\ \bibinfo {pages} {195116} (\bibinfo {year}
  {2016})}\BibitemShut {NoStop}%
\bibitem [{\citenamefont {Peterseim}\ \emph {et~al.}(2016)\citenamefont
  {Peterseim}, \citenamefont {Ivek}, \citenamefont {Schweitzer},\ and\
  \citenamefont {Dressel}}]{Peterseim16}%
  \BibitemOpen
  \bibfield  {author} {\bibinfo {author} {\bibfnamefont {T.}~\bibnamefont
  {Peterseim}}, \bibinfo {author} {\bibfnamefont {T.}~\bibnamefont {Ivek}},
  \bibinfo {author} {\bibfnamefont {D.}~\bibnamefont {Schweitzer}}, \ and\
  \bibinfo {author} {\bibfnamefont {M.}~\bibnamefont {Dressel}},\ }\href
  {\doibase 10.1103/PhysRevB.93.245133} {\bibfield  {journal} {\bibinfo
  {journal} {Phys. Rev. B}\ }\textbf {\bibinfo {volume} {93}},\ \bibinfo
  {pages} {245133} (\bibinfo {year} {2016})}\BibitemShut {NoStop}%
\bibitem [{\citenamefont {{Hirata}}\ \emph {et~al.}(2016)\citenamefont
  {{Hirata}}, \citenamefont {{Ishikawa}}, \citenamefont {{Miyagawa}},
  \citenamefont {{Tamura}}, \citenamefont {{Berthier}}, \citenamefont
  {{Basko}}, \citenamefont {{Kobayashi}}, \citenamefont {{Matsuno}},\ and\
  \citenamefont {{Kanoda}}}]{Hirata16}%
  \BibitemOpen
  \bibfield  {author} {\bibinfo {author} {\bibfnamefont {M.}~\bibnamefont
  {{Hirata}}}, \bibinfo {author} {\bibfnamefont {K.}~\bibnamefont
  {{Ishikawa}}}, \bibinfo {author} {\bibfnamefont {K.}~\bibnamefont
  {{Miyagawa}}}, \bibinfo {author} {\bibfnamefont {M.}~\bibnamefont
  {{Tamura}}}, \bibinfo {author} {\bibfnamefont {C.}~\bibnamefont
  {{Berthier}}}, \bibinfo {author} {\bibfnamefont {D.}~\bibnamefont {{Basko}}},
  \bibinfo {author} {\bibfnamefont {A.}~\bibnamefont {{Kobayashi}}}, \bibinfo
  {author} {\bibfnamefont {G.}~\bibnamefont {{Matsuno}}}, \ and\ \bibinfo
  {author} {\bibfnamefont {K.}~\bibnamefont {{Kanoda}}},\ }\href {\doibase
  10.1038/ncomms12666} {\bibfield  {journal} {\bibinfo  {journal} {Nature
  Communications}\ }\textbf {\bibinfo {volume} {7}},\ \bibinfo {eid} {12666}
  (\bibinfo {year} {2016})},\ \Eprint {http://arxiv.org/abs/1607.07142}
  {arXiv:1607.07142 [cond-mat.str-el]} \BibitemShut {NoStop}%
\bibitem [{\citenamefont {Hirata}\ \emph {et~al.}(2017)\citenamefont {Hirata},
  \citenamefont {Ishikawa}, \citenamefont {Matsuno}, \citenamefont {Kobayashi},
  \citenamefont {Miyagawa}, \citenamefont {Tamura}, \citenamefont {Berthier},\
  and\ \citenamefont {Kanoda}}]{Hirata17}%
  \BibitemOpen
  \bibfield  {author} {\bibinfo {author} {\bibfnamefont {M.}~\bibnamefont
  {Hirata}}, \bibinfo {author} {\bibfnamefont {K.}~\bibnamefont {Ishikawa}},
  \bibinfo {author} {\bibfnamefont {G.}~\bibnamefont {Matsuno}}, \bibinfo
  {author} {\bibfnamefont {A.}~\bibnamefont {Kobayashi}}, \bibinfo {author}
  {\bibfnamefont {K.}~\bibnamefont {Miyagawa}}, \bibinfo {author}
  {\bibfnamefont {M.}~\bibnamefont {Tamura}}, \bibinfo {author} {\bibfnamefont
  {C.}~\bibnamefont {Berthier}}, \ and\ \bibinfo {author} {\bibfnamefont
  {K.}~\bibnamefont {Kanoda}},\ }\href {\doibase 10.1126/science.aan5351}
  {\bibfield  {journal} {\bibinfo  {journal} {Science}\ }\textbf {\bibinfo
  {volume} {358}},\ \bibinfo {pages} {1403} (\bibinfo {year} {2017})},\ \Eprint
  {http://arxiv.org/abs/http://science.sciencemag.org/content/358/6369/1403.full.pdf}
  {http://science.sciencemag.org/content/358/6369/1403.full.pdf} \BibitemShut
  {NoStop}%
\bibitem [{\citenamefont {Kotov}\ \emph {et~al.}(2012)\citenamefont {Kotov},
  \citenamefont {Uchoa}, \citenamefont {Pereira}, \citenamefont {Guinea},\ and\
  \citenamefont {Castro~Neto}}]{Kotov12}%
  \BibitemOpen
  \bibfield  {author} {\bibinfo {author} {\bibfnamefont {V.~N.}\ \bibnamefont
  {Kotov}}, \bibinfo {author} {\bibfnamefont {B.}~\bibnamefont {Uchoa}},
  \bibinfo {author} {\bibfnamefont {V.~M.}\ \bibnamefont {Pereira}}, \bibinfo
  {author} {\bibfnamefont {F.}~\bibnamefont {Guinea}}, \ and\ \bibinfo {author}
  {\bibfnamefont {A.~H.}\ \bibnamefont {Castro~Neto}},\ }\href {\doibase
  10.1103/RevModPhys.84.1067} {\bibfield  {journal} {\bibinfo  {journal} {Rev.
  Mod. Phys.}\ }\textbf {\bibinfo {volume} {84}},\ \bibinfo {pages} {1067}
  (\bibinfo {year} {2012})}\BibitemShut {NoStop}%
\bibitem [{\citenamefont {Khveshchenko}(2009)}]{Knveshchenko09}%
  \BibitemOpen
  \bibfield  {author} {\bibinfo {author} {\bibfnamefont {D.~V.}\ \bibnamefont
  {Khveshchenko}},\ }\href {http://stacks.iop.org/0953-8984/21/i=7/a=075303}
  {\bibfield  {journal} {\bibinfo  {journal} {Journal of Physics: Condensed
  Matter}\ }\textbf {\bibinfo {volume} {21}},\ \bibinfo {pages} {075303}
  (\bibinfo {year} {2009})}\BibitemShut {NoStop}%
\bibitem [{\citenamefont {{Zhang}}\ \emph {et~al.}(2008)\citenamefont
  {{Zhang}}, \citenamefont {{Brar}}, \citenamefont {{Wang}}, \citenamefont
  {{Girit}}, \citenamefont {{Yayon}}, \citenamefont {{Panlasigui}},
  \citenamefont {{Zettl}},\ and\ \citenamefont {{Crommie}}}]{Zhang08}%
  \BibitemOpen
  \bibfield  {author} {\bibinfo {author} {\bibfnamefont {Y.}~\bibnamefont
  {{Zhang}}}, \bibinfo {author} {\bibfnamefont {V.~W.}\ \bibnamefont {{Brar}}},
  \bibinfo {author} {\bibfnamefont {F.}~\bibnamefont {{Wang}}}, \bibinfo
  {author} {\bibfnamefont {C.}~\bibnamefont {{Girit}}}, \bibinfo {author}
  {\bibfnamefont {Y.}~\bibnamefont {{Yayon}}}, \bibinfo {author} {\bibfnamefont
  {M.}~\bibnamefont {{Panlasigui}}}, \bibinfo {author} {\bibfnamefont
  {A.}~\bibnamefont {{Zettl}}}, \ and\ \bibinfo {author} {\bibfnamefont
  {M.~F.}\ \bibnamefont {{Crommie}}},\ }\href {\doibase 10.1038/nphys1022}
  {\bibfield  {journal} {\bibinfo  {journal} {Nature Physics}\ }\textbf
  {\bibinfo {volume} {4}},\ \bibinfo {pages} {627} (\bibinfo {year} {2008})},\
  \Eprint {http://arxiv.org/abs/0802.4315} {arXiv:0802.4315
  [cond-mat.mes-hall]} \BibitemShut {NoStop}%
\bibitem [{\citenamefont {Monteverde}\ \emph {et~al.}(2013)\citenamefont
  {Monteverde}, \citenamefont {Goerbig}, \citenamefont {Auban-Senzier},
  \citenamefont {Navarin}, \citenamefont {Henck}, \citenamefont {Pasquier},
  \citenamefont {M\'ezi\`ere},\ and\ \citenamefont {Batail}}]{Monteverde13}%
  \BibitemOpen
  \bibfield  {author} {\bibinfo {author} {\bibfnamefont {M.}~\bibnamefont
  {Monteverde}}, \bibinfo {author} {\bibfnamefont {M.~O.}\ \bibnamefont
  {Goerbig}}, \bibinfo {author} {\bibfnamefont {P.}~\bibnamefont
  {Auban-Senzier}}, \bibinfo {author} {\bibfnamefont {F.}~\bibnamefont
  {Navarin}}, \bibinfo {author} {\bibfnamefont {H.}~\bibnamefont {Henck}},
  \bibinfo {author} {\bibfnamefont {C.~R.}\ \bibnamefont {Pasquier}}, \bibinfo
  {author} {\bibfnamefont {C.}~\bibnamefont {M\'ezi\`ere}}, \ and\ \bibinfo
  {author} {\bibfnamefont {P.}~\bibnamefont {Batail}},\ }\href {\doibase
  10.1103/PhysRevB.87.245110} {\bibfield  {journal} {\bibinfo  {journal} {Phys.
  Rev. B}\ }\textbf {\bibinfo {volume} {87}},\ \bibinfo {pages} {245110}
  (\bibinfo {year} {2013})}\BibitemShut {NoStop}%
\bibitem [{\citenamefont {Mak}\ \emph {et~al.}(2008)\citenamefont {Mak},
  \citenamefont {Sfeir}, \citenamefont {Wu}, \citenamefont {Lui}, \citenamefont
  {Misewich},\ and\ \citenamefont {Heinz}}]{Mak08}%
  \BibitemOpen
  \bibfield  {author} {\bibinfo {author} {\bibfnamefont {K.~F.}\ \bibnamefont
  {Mak}}, \bibinfo {author} {\bibfnamefont {M.~Y.}\ \bibnamefont {Sfeir}},
  \bibinfo {author} {\bibfnamefont {Y.}~\bibnamefont {Wu}}, \bibinfo {author}
  {\bibfnamefont {C.~H.}\ \bibnamefont {Lui}}, \bibinfo {author} {\bibfnamefont
  {J.~A.}\ \bibnamefont {Misewich}}, \ and\ \bibinfo {author} {\bibfnamefont
  {T.~F.}\ \bibnamefont {Heinz}},\ }\href {\doibase
  10.1103/PhysRevLett.101.196405} {\bibfield  {journal} {\bibinfo  {journal}
  {Phys. Rev. Lett.}\ }\textbf {\bibinfo {volume} {101}},\ \bibinfo {pages}
  {196405} (\bibinfo {year} {2008})}\BibitemShut {NoStop}%
\bibitem [{\citenamefont {Kuzmenko}\ \emph {et~al.}(2008)\citenamefont
  {Kuzmenko}, \citenamefont {van Heumen}, \citenamefont {Carbone},\ and\
  \citenamefont {van~der Marel}}]{Kuzmenko08}%
  \BibitemOpen
  \bibfield  {author} {\bibinfo {author} {\bibfnamefont {A.~B.}\ \bibnamefont
  {Kuzmenko}}, \bibinfo {author} {\bibfnamefont {E.}~\bibnamefont {van
  Heumen}}, \bibinfo {author} {\bibfnamefont {F.}~\bibnamefont {Carbone}}, \
  and\ \bibinfo {author} {\bibfnamefont {D.}~\bibnamefont {van~der Marel}},\
  }\href {\doibase 10.1103/PhysRevLett.100.117401} {\bibfield  {journal}
  {\bibinfo  {journal} {Phys. Rev. Lett.}\ }\textbf {\bibinfo {volume} {100}},\
  \bibinfo {pages} {117401} (\bibinfo {year} {2008})}\BibitemShut {NoStop}%
\bibitem [{\citenamefont {Keller}\ and\ \citenamefont
  {Holzapfel}(1977)}]{Keller77}%
  \BibitemOpen
  \bibfield  {author} {\bibinfo {author} {\bibfnamefont {R.}~\bibnamefont
  {Keller}}\ and\ \bibinfo {author} {\bibfnamefont {W.~B.}\ \bibnamefont
  {Holzapfel}},\ }\href {\doibase 10.1063/1.1135065} {\bibfield  {journal}
  {\bibinfo  {journal} {Review of Scientific Instruments}\ }\textbf {\bibinfo
  {volume} {48}},\ \bibinfo {pages} {517} (\bibinfo {year} {1977})},\ \Eprint
  {http://arxiv.org/abs/http://dx.doi.org/10.1063/1.1135065}
  {http://dx.doi.org/10.1063/1.1135065} \BibitemShut {NoStop}%
\bibitem [{\citenamefont {Kuntscher}\ \emph {et~al.}(2014)\citenamefont
  {Kuntscher}, \citenamefont {Huber},\ and\ \citenamefont
  {H\"ucker}}]{Kuntscher14}%
  \BibitemOpen
  \bibfield  {author} {\bibinfo {author} {\bibfnamefont {C.~A.}\ \bibnamefont
  {Kuntscher}}, \bibinfo {author} {\bibfnamefont {A.}~\bibnamefont {Huber}}, \
  and\ \bibinfo {author} {\bibfnamefont {M.}~\bibnamefont {H\"ucker}},\ }\href
  {\doibase 10.1103/PhysRevB.89.134510} {\bibfield  {journal} {\bibinfo
  {journal} {Phys. Rev. B}\ }\textbf {\bibinfo {volume} {89}},\ \bibinfo
  {pages} {134510} (\bibinfo {year} {2014})}\BibitemShut {NoStop}%
\bibitem [{\citenamefont {Mao}\ \emph {et~al.}(1986)\citenamefont {Mao},
  \citenamefont {Xu},\ and\ \citenamefont {Bell}}]{Mao86}%
  \BibitemOpen
  \bibfield  {author} {\bibinfo {author} {\bibfnamefont {H.~K.}\ \bibnamefont
  {Mao}}, \bibinfo {author} {\bibfnamefont {J.}~\bibnamefont {Xu}}, \ and\
  \bibinfo {author} {\bibfnamefont {P.~M.}\ \bibnamefont {Bell}},\ }\href
  {\doibase 10.1029/JB091iB05p04673} {\bibfield  {journal} {\bibinfo  {journal}
  {J. Geophys. Res. Solid Earth}\ }\textbf {\bibinfo {volume} {91}},\ \bibinfo
  {pages} {4673} (\bibinfo {year} {1986})}\BibitemShut {NoStop}%
\bibitem [{\citenamefont {Pashkin}\ \emph {et~al.}(2006)\citenamefont
  {Pashkin}, \citenamefont {Dressel},\ and\ \citenamefont
  {Kuntscher}}]{Pashkin06}%
  \BibitemOpen
  \bibfield  {author} {\bibinfo {author} {\bibfnamefont {A.}~\bibnamefont
  {Pashkin}}, \bibinfo {author} {\bibfnamefont {M.}~\bibnamefont {Dressel}}, \
  and\ \bibinfo {author} {\bibfnamefont {C.~A.}\ \bibnamefont {Kuntscher}},\
  }\href {\doibase 10.1103/PhysRevB.74.165118} {\bibfield  {journal} {\bibinfo
  {journal} {Phys. Rev. B}\ }\textbf {\bibinfo {volume} {74}},\ \bibinfo
  {pages} {165118} (\bibinfo {year} {2006})}\BibitemShut {NoStop}%
\bibitem [{SM()}]{SM}%
  \BibitemOpen
  \href@noop {} {}\bibinfo {note} {See the Supplemental Materials for details
  on the crystals, samples preparation, and measurements.}\BibitemShut {Stop}%
\bibitem [{\citenamefont {{M. Dressel}}\ \emph {et~al.}(1994)\citenamefont {{M.
  Dressel}}, \citenamefont {{G. Grüner}}, \citenamefont {{J.P. Pouget}},
  \citenamefont {{A. Breining}},\ and\ \citenamefont {{D.
  Schweitzer}}}]{Dressel94}%
  \BibitemOpen
  \bibfield  {author} {\bibinfo {author} {\bibnamefont {{M. Dressel}}},
  \bibinfo {author} {\bibnamefont {{G. Grüner}}}, \bibinfo {author}
  {\bibnamefont {{J.P. Pouget}}}, \bibinfo {author} {\bibnamefont {{A.
  Breining}}}, \ and\ \bibinfo {author} {\bibnamefont {{D. Schweitzer}}},\
  }\href {\doibase 10.1051/jp1:1994162} {\bibfield  {journal} {\bibinfo
  {journal} {J. Phys. I France}\ }\textbf {\bibinfo {volume} {4}},\ \bibinfo
  {pages} {579} (\bibinfo {year} {1994})}\BibitemShut {NoStop}%
\bibitem [{\citenamefont {Dressel}\ and\ \citenamefont
  {Drichko}(2004)}]{Dressel04}%
  \BibitemOpen
  \bibfield  {author} {\bibinfo {author} {\bibfnamefont {M.}~\bibnamefont
  {Dressel}}\ and\ \bibinfo {author} {\bibfnamefont {N.}~\bibnamefont
  {Drichko}},\ }\href {\doibase 10.1021/cr030642f} {\bibfield  {journal}
  {\bibinfo  {journal} {Chemical Reviews}\ }\textbf {\bibinfo {volume} {104}},\
  \bibinfo {pages} {5689} (\bibinfo {year} {2004})},\ \bibinfo {note} {pMID:
  15535665},\ \Eprint
  {http://arxiv.org/abs/http://dx.doi.org/10.1021/cr030642f}
  {http://dx.doi.org/10.1021/cr030642f} \BibitemShut {NoStop}%
\bibitem [{\citenamefont {Basov}\ \emph {et~al.}(2011)\citenamefont {Basov},
  \citenamefont {Averitt}, \citenamefont {van~der Marel}, \citenamefont
  {Dressel},\ and\ \citenamefont {Haule}}]{Basov11}%
  \BibitemOpen
  \bibfield  {author} {\bibinfo {author} {\bibfnamefont {D.~N.}\ \bibnamefont
  {Basov}}, \bibinfo {author} {\bibfnamefont {R.~D.}\ \bibnamefont {Averitt}},
  \bibinfo {author} {\bibfnamefont {D.}~\bibnamefont {van~der Marel}}, \bibinfo
  {author} {\bibfnamefont {M.}~\bibnamefont {Dressel}}, \ and\ \bibinfo
  {author} {\bibfnamefont {K.}~\bibnamefont {Haule}},\ }\href {\doibase
  10.1103/RevModPhys.83.471} {\bibfield  {journal} {\bibinfo  {journal} {Rev.
  Mod. Phys.}\ }\textbf {\bibinfo {volume} {83}},\ \bibinfo {pages} {471}
  (\bibinfo {year} {2011})}\BibitemShut {NoStop}%
\bibitem [{\citenamefont {Drichko}\ \emph {et~al.}(2006)\citenamefont
  {Drichko}, \citenamefont {Dressel}, \citenamefont {Kuntscher}, \citenamefont
  {Pashkin}, \citenamefont {Greco}, \citenamefont {Merino},\ and\ \citenamefont
  {Schlueter}}]{Drichko06}%
  \BibitemOpen
  \bibfield  {author} {\bibinfo {author} {\bibfnamefont {N.}~\bibnamefont
  {Drichko}}, \bibinfo {author} {\bibfnamefont {M.}~\bibnamefont {Dressel}},
  \bibinfo {author} {\bibfnamefont {C.~A.}\ \bibnamefont {Kuntscher}}, \bibinfo
  {author} {\bibfnamefont {A.}~\bibnamefont {Pashkin}}, \bibinfo {author}
  {\bibfnamefont {A.}~\bibnamefont {Greco}}, \bibinfo {author} {\bibfnamefont
  {J.}~\bibnamefont {Merino}}, \ and\ \bibinfo {author} {\bibfnamefont
  {J.}~\bibnamefont {Schlueter}},\ }\href {\doibase 10.1103/PhysRevB.74.235121}
  {\bibfield  {journal} {\bibinfo  {journal} {Phys. Rev. B}\ }\textbf {\bibinfo
  {volume} {74}},\ \bibinfo {pages} {235121} (\bibinfo {year}
  {2006})}\BibitemShut {NoStop}%
\bibitem [{\citenamefont {Rozenberg}\ \emph {et~al.}(1995)\citenamefont
  {Rozenberg}, \citenamefont {Kotliar}, \citenamefont {Kajueter}, \citenamefont
  {Thomas}, \citenamefont {Rapkine}, \citenamefont {Honig},\ and\ \citenamefont
  {Metcalf}}]{Rozenberg95}%
  \BibitemOpen
  \bibfield  {author} {\bibinfo {author} {\bibfnamefont {M.}~\bibnamefont
  {Rozenberg}}, \bibinfo {author} {\bibfnamefont {G.}~\bibnamefont {Kotliar}},
  \bibinfo {author} {\bibfnamefont {H.}~\bibnamefont {Kajueter}}, \bibinfo
  {author} {\bibfnamefont {G.}~\bibnamefont {Thomas}}, \bibinfo {author}
  {\bibfnamefont {D.}~\bibnamefont {Rapkine}}, \bibinfo {author} {\bibfnamefont
  {J.}~\bibnamefont {Honig}}, \ and\ \bibinfo {author} {\bibfnamefont
  {P.}~\bibnamefont {Metcalf}},\ }\href@noop {} {\bibfield  {journal} {\bibinfo
   {journal} {Phys. Rev. Lett.}\ }\textbf {\bibinfo {volume} {75}},\ \bibinfo
  {pages} {105} (\bibinfo {year} {1995})}\BibitemShut {NoStop}%
\bibitem [{\citenamefont {Kobayashi}\ \emph {et~al.}(2009)\citenamefont
  {Kobayashi}, \citenamefont {Katayama},\ and\ \citenamefont
  {Suzumura}}]{Kobayashi09}%
  \BibitemOpen
  \bibfield  {author} {\bibinfo {author} {\bibfnamefont {A.}~\bibnamefont
  {Kobayashi}}, \bibinfo {author} {\bibfnamefont {S.}~\bibnamefont {Katayama}},
  \ and\ \bibinfo {author} {\bibfnamefont {Y.}~\bibnamefont {Suzumura}},\
  }\href {http://stacks.iop.org/1468-6996/10/i=2/a=024309} {\bibfield
  {journal} {\bibinfo  {journal} {Science and Technology of Advanced
  Materials}\ }\textbf {\bibinfo {volume} {10}},\ \bibinfo {pages} {024309}
  (\bibinfo {year} {2009})}\BibitemShut {NoStop}%
\bibitem [{\citenamefont {Kondo}\ \emph {et~al.}(2009)\citenamefont {Kondo},
  \citenamefont {Kagoshima}, \citenamefont {Tajima},\ and\ \citenamefont
  {Kato}}]{Kondo09}%
  \BibitemOpen
  \bibfield  {author} {\bibinfo {author} {\bibfnamefont {R.}~\bibnamefont
  {Kondo}}, \bibinfo {author} {\bibfnamefont {S.}~\bibnamefont {Kagoshima}},
  \bibinfo {author} {\bibfnamefont {N.}~\bibnamefont {Tajima}}, \ and\ \bibinfo
  {author} {\bibfnamefont {R.}~\bibnamefont {Kato}},\ }\href {\doibase
  10.1143/JPSJ.78.114714} {\bibfield  {journal} {\bibinfo  {journal} {J. Phys.
  Soc. Jpn.}\ }\textbf {\bibinfo {volume} {78}},\ \bibinfo {pages} {114714}
  (\bibinfo {year} {2009})},\ \Eprint
  {http://arxiv.org/abs/http://dx.doi.org/10.1143/JPSJ.78.114714}
  {http://dx.doi.org/10.1143/JPSJ.78.114714} \BibitemShut {NoStop}%
\bibitem [{\citenamefont {Perucchi}\ \emph {et~al.}(2004)\citenamefont
  {Perucchi}, \citenamefont {Degiorgi},\ and\ \citenamefont
  {Thorne}}]{Perucchi04}%
  \BibitemOpen
  \bibfield  {author} {\bibinfo {author} {\bibfnamefont {A.}~\bibnamefont
  {Perucchi}}, \bibinfo {author} {\bibfnamefont {L.}~\bibnamefont {Degiorgi}},
  \ and\ \bibinfo {author} {\bibfnamefont {R.~E.}\ \bibnamefont {Thorne}},\
  }\href {\doibase 10.1103/PhysRevB.69.195114} {\bibfield  {journal} {\bibinfo
  {journal} {Phys. Rev. B}\ }\textbf {\bibinfo {volume} {69}},\ \bibinfo
  {pages} {195114} (\bibinfo {year} {2004})}\BibitemShut {NoStop}%
\bibitem [{\citenamefont {Bari\ifmmode \check{s}\else
  \v{s}\fi{}i\ifmmode~\acute{c}\else \'{c}\fi{}}\ \emph
  {et~al.}(2010)\citenamefont {Bari\ifmmode \check{s}\else
  \v{s}\fi{}i\ifmmode~\acute{c}\else \'{c}\fi{}}, \citenamefont {Wu},
  \citenamefont {Dressel}, \citenamefont {Li}, \citenamefont {Cao},\ and\
  \citenamefont {Xu}}]{Barisic10}%
  \BibitemOpen
  \bibfield  {author} {\bibinfo {author} {\bibfnamefont {N.}~\bibnamefont
  {Bari\ifmmode \check{s}\else \v{s}\fi{}i\ifmmode~\acute{c}\else \'{c}\fi{}}},
  \bibinfo {author} {\bibfnamefont {D.}~\bibnamefont {Wu}}, \bibinfo {author}
  {\bibfnamefont {M.}~\bibnamefont {Dressel}}, \bibinfo {author} {\bibfnamefont
  {L.~J.}\ \bibnamefont {Li}}, \bibinfo {author} {\bibfnamefont {G.~H.}\
  \bibnamefont {Cao}}, \ and\ \bibinfo {author} {\bibfnamefont {Z.~A.}\
  \bibnamefont {Xu}},\ }\href {\doibase 10.1103/PhysRevB.82.054518} {\bibfield
  {journal} {\bibinfo  {journal} {Phys. Rev. B}\ }\textbf {\bibinfo {volume}
  {82}},\ \bibinfo {pages} {054518} (\bibinfo {year} {2010})}\BibitemShut
  {NoStop}%
\bibitem [{\citenamefont {Dumm}\ \emph {et~al.}(2009)\citenamefont {Dumm},
  \citenamefont {Faltermeier}, \citenamefont {Drichko}, \citenamefont
  {Dressel}, \citenamefont {M\'ezi\`ere},\ and\ \citenamefont
  {Batail}}]{Dumm09}%
  \BibitemOpen
  \bibfield  {author} {\bibinfo {author} {\bibfnamefont {M.}~\bibnamefont
  {Dumm}}, \bibinfo {author} {\bibfnamefont {D.}~\bibnamefont {Faltermeier}},
  \bibinfo {author} {\bibfnamefont {N.}~\bibnamefont {Drichko}}, \bibinfo
  {author} {\bibfnamefont {M.}~\bibnamefont {Dressel}}, \bibinfo {author}
  {\bibfnamefont {C.}~\bibnamefont {M\'ezi\`ere}}, \ and\ \bibinfo {author}
  {\bibfnamefont {P.}~\bibnamefont {Batail}},\ }\href {\doibase
  10.1103/PhysRevB.79.195106} {\bibfield  {journal} {\bibinfo  {journal} {Phys.
  Rev. B}\ }\textbf {\bibinfo {volume} {79}},\ \bibinfo {pages} {195106}
  (\bibinfo {year} {2009})}\BibitemShut {NoStop}%
\bibitem [{\citenamefont {Pustogow}\ \emph {et~al.}()\citenamefont {Pustogow},
  \citenamefont {Bories}, \citenamefont {Löhle}, \citenamefont {Rösslhuber},
  \citenamefont {Zhukova}, \citenamefont {Gorshunov}, \citenamefont {Tomić},
  \citenamefont {Schlueter}, \citenamefont {Hübner}, \citenamefont
  {Hiramatsu}, \citenamefont {Yoshida}, \citenamefont {Saito}, \citenamefont
  {Kato}, \citenamefont {Lee}, \citenamefont {Dobrosavljević}, \citenamefont
  {Fratini},\ and\ \citenamefont {Dressel}}]{Pustogow17}%
  \BibitemOpen
  \bibfield  {author} {\bibinfo {author} {\bibfnamefont {A.}~\bibnamefont
  {Pustogow}}, \bibinfo {author} {\bibfnamefont {M.}~\bibnamefont {Bories}},
  \bibinfo {author} {\bibfnamefont {A.}~\bibnamefont {Löhle}}, \bibinfo
  {author} {\bibfnamefont {R.}~\bibnamefont {Rösslhuber}}, \bibinfo {author}
  {\bibfnamefont {E.}~\bibnamefont {Zhukova}}, \bibinfo {author} {\bibfnamefont
  {B.}~\bibnamefont {Gorshunov}}, \bibinfo {author} {\bibfnamefont
  {S.}~\bibnamefont {Tomić}}, \bibinfo {author} {\bibfnamefont
  {J.}~\bibnamefont {Schlueter}}, \bibinfo {author} {\bibfnamefont
  {R.}~\bibnamefont {Hübner}}, \bibinfo {author} {\bibfnamefont
  {T.}~\bibnamefont {Hiramatsu}}, \bibinfo {author} {\bibfnamefont
  {Y.}~\bibnamefont {Yoshida}}, \bibinfo {author} {\bibfnamefont
  {G.}~\bibnamefont {Saito}}, \bibinfo {author} {\bibfnamefont
  {R.}~\bibnamefont {Kato}}, \bibinfo {author} {\bibfnamefont {T.-H.}\
  \bibnamefont {Lee}}, \bibinfo {author} {\bibfnamefont {V.}~\bibnamefont
  {Dobrosavljević}}, \bibinfo {author} {\bibfnamefont {S.}~\bibnamefont
  {Fratini}}, \ and\ \bibinfo {author} {\bibfnamefont {M.}~\bibnamefont
  {Dressel}},\ }\href@noop {} {\bibinfo  {journal} {ArXiv: 1710.07241}\
  }\BibitemShut {NoStop}%
\bibitem [{\citenamefont {Suzumura}\ \emph {et~al.}(2014)\citenamefont
  {Suzumura}, \citenamefont {Proskurin},\ and\ \citenamefont
  {Ogata}}]{Suzumura14}%
  \BibitemOpen
\bibfield  {journal} {  }\bibfield  {author} {\bibinfo {author} {\bibfnamefont
  {Y.}~\bibnamefont {Suzumura}}, \bibinfo {author} {\bibfnamefont
  {I.}~\bibnamefont {Proskurin}}, \ and\ \bibinfo {author} {\bibfnamefont
  {M.}~\bibnamefont {Ogata}},\ }\href {\doibase 10.7566/JPSJ.83.094705}
  {\bibfield  {journal} {\bibinfo  {journal} {J. Phys. Soc. Jpn.}\ }\textbf
  {\bibinfo {volume} {83}},\ \bibinfo {pages} {094705} (\bibinfo {year}
  {2014})},\ \Eprint
  {http://arxiv.org/abs/http://dx.doi.org/10.7566/JPSJ.83.094705}
  {http://dx.doi.org/10.7566/JPSJ.83.094705} \BibitemShut {NoStop}%
\bibitem [{\citenamefont {Dressel}\ and\ \citenamefont
  {Gr{\"u}ner}(2002)}]{DresselGruner02}%
  \BibitemOpen
  \bibfield  {author} {\bibinfo {author} {\bibfnamefont {M.}~\bibnamefont
  {Dressel}}\ and\ \bibinfo {author} {\bibfnamefont {G.}~\bibnamefont
  {Gr{\"u}ner}},\ }\href@noop {} {\emph {\bibinfo {title} {Electrodynamics of
  Solids}}}\ (\bibinfo  {publisher} {Cambridge University Press},\ \bibinfo
  {address} {Cambridge},\ \bibinfo {year} {2002})\BibitemShut {NoStop}%
\bibitem [{\citenamefont {Kishigi}\ and\ \citenamefont
  {Hasegawa}(2017)}]{Kishigi17}%
  \BibitemOpen
  \bibfield  {author} {\bibinfo {author} {\bibfnamefont {K.}~\bibnamefont
  {Kishigi}}\ and\ \bibinfo {author} {\bibfnamefont {Y.}~\bibnamefont
  {Hasegawa}},\ }\href {\doibase 10.1103/PhysRevB.96.085430} {\bibfield
  {journal} {\bibinfo  {journal} {Phys. Rev. B}\ }\textbf {\bibinfo {volume}
  {96}},\ \bibinfo {pages} {085430} (\bibinfo {year} {2017})}\BibitemShut
  {NoStop}%
\end{thebibliography}%

\newpage
\setcounter{figure}{0}
\renewcommand{\thefigure}{S\arabic{figure}}
\renewcommand{\citenumfont}[1]{S#1}

\onecolumngrid
\appendix

\begin{center}
\textbf{Supplementary information for\\ ``Electronic Correlations Among the Dirac Electrons in $\alpha$-(BEDT-TTF)$_3$I$_3$\\ Unveiled by High-Pressure Optical Spectroscopy"}
\end{center}
\centerline{Weiwu Li, Ece Uykur, Christine A. Kuntscher, Martin Dressel}

\section{Experimental}

The temperature-dependent high pressure reflectivity measurements were performed on \aeti\ single crystals with the typical size of 300~$\mu$m~$\times$~300~$\mu$m~$\times$~60~$\mu$m. Two pieces of flat, high-quality crystals, cut from the same big piece, were used for the far-infrared and the middle-infrared measurements. $E\mid\mid ab$-plane measurement configuration has been chosen and measurements have been conducted without a polarizer.

Samples were placed inside a diamond anvil cell (DAC) with a culet diameter of 900~$\mu$m with ruby spheres as the high pressure manometers. Finely grounded CsI powder was employed as a quasihydrostatic pressure transmitting medium. Standart ruby luminescence technique [S1] has been used to determine the pressure inside the cell at each temperature.

Temperature-dependent reflectivity measurements were performed from $\sim$~100 to 8000~\cm\ between 6 and 300~K using a homebuilt IR-microscope setup that is coupled to a Bruker Vertex 80v Fourier transform infrared spectrometer [S2]. The measured pressure range extends up to $\sim$~4~GPa. The reflectivity spectra were measured at the sample diamond interface, using the CuBe gasket inside the DAC as reference. Afterwards, the intensity of the measured reflectivity was normalized by the intensity reflected from the CuBe gasket at the gasket-diamond interface.

Optical conductivity spectra have been obtained with Kramers-Kronig analysis from reflectivity measurements [S3]. Between 1700 and 2800~\cm\ the multiphonon absorption of the diamond anvil affects the measured spectra, therefore, in this energy range a linear extrapolation have been used for further analysis. Despite the bad metallicity in this system, the reflectivity shows a clear upturn. Therefore, in the low energy range Hagen-Rubens extrapolation has been chosen. For the high energy range a flat conductivity followed by $\omega^{-4}$ free carrier approximation have been utilized. 

\begin{figure*}[h]%
\centering
\includegraphics[scale=1.2]{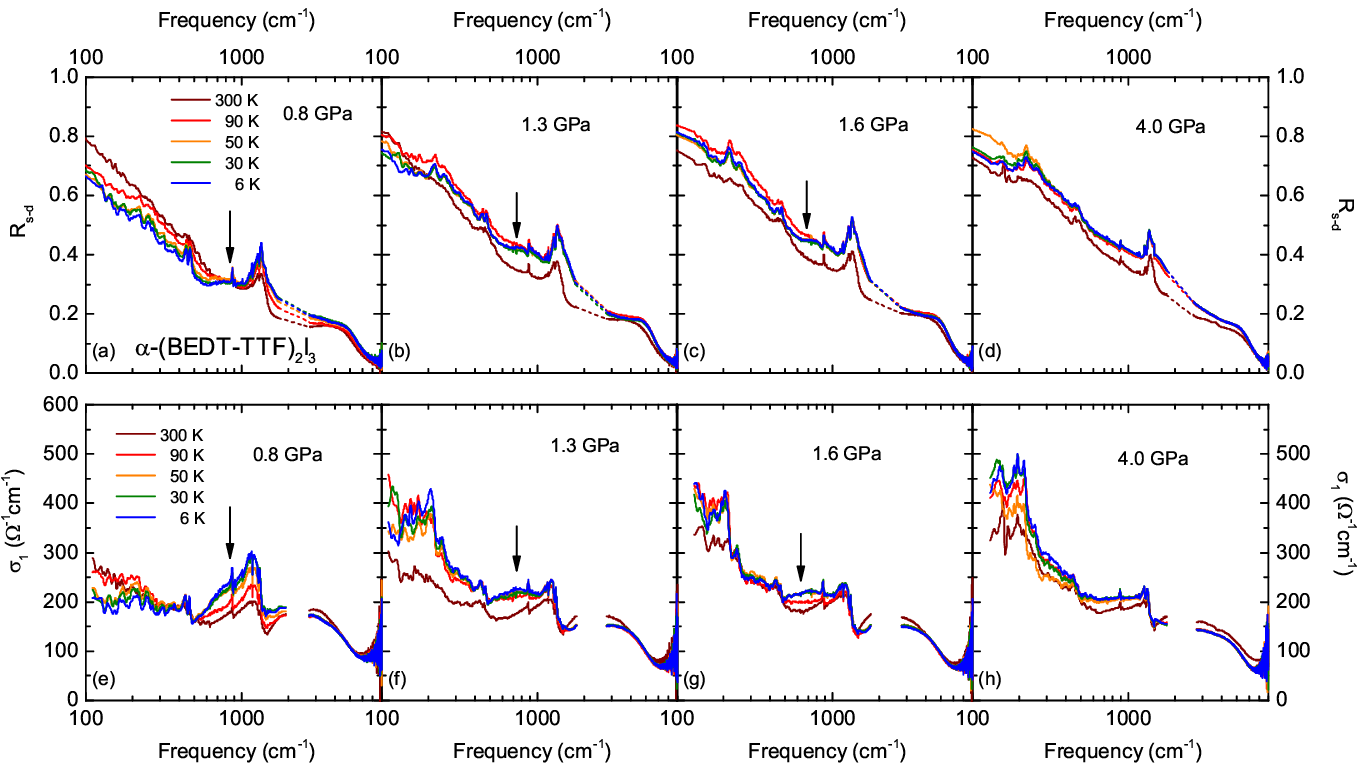}%
\caption{Temperature and pressure-dependent reflectivity (a-d) and corresponding optical conductivity (e-h) spectra of \aeti\ . A linear extrapolation have been used in the multiphonon diamond absorption region shown with dashed curves (a-d). Arrows mark the spectral weight transfer due to pseudogap behavior.}%
\label{RefCon}%
\end{figure*}

\section{Reflectivity and Optical Conductivity}

Fig.~\ref{RefCon} present the temperature-dependent reflectivity (a-d) (as-measured) and the calculated optical conductivity spectra (e-h) for various pressures.

In Figure~\ref{RefCon}(a-c), one can see the suppression of the low frequency reflectivity below a certain temperature, while a clear bending of the spectra is also marked with arrows, which give rise the absorption-like feature in the corresponding optical conductivity spectra (Fig.~\ref{RefCon}(e-g)). At 4.0~GPa this behavior is suppressed and/or shift to the very low energy range that we cannot resolve it anymore. This absorption feature has been discussed in the mansucript in terms of pseudogap behavior, which is also can be clearly seen in the as-measured spectra. While the origin and the behavior of the vibration modes are out of scope of this study, we preferred to subtract them to make the discussion easier. 

\begin{figure}[h]%
\centering
\includegraphics[scale=0.75]{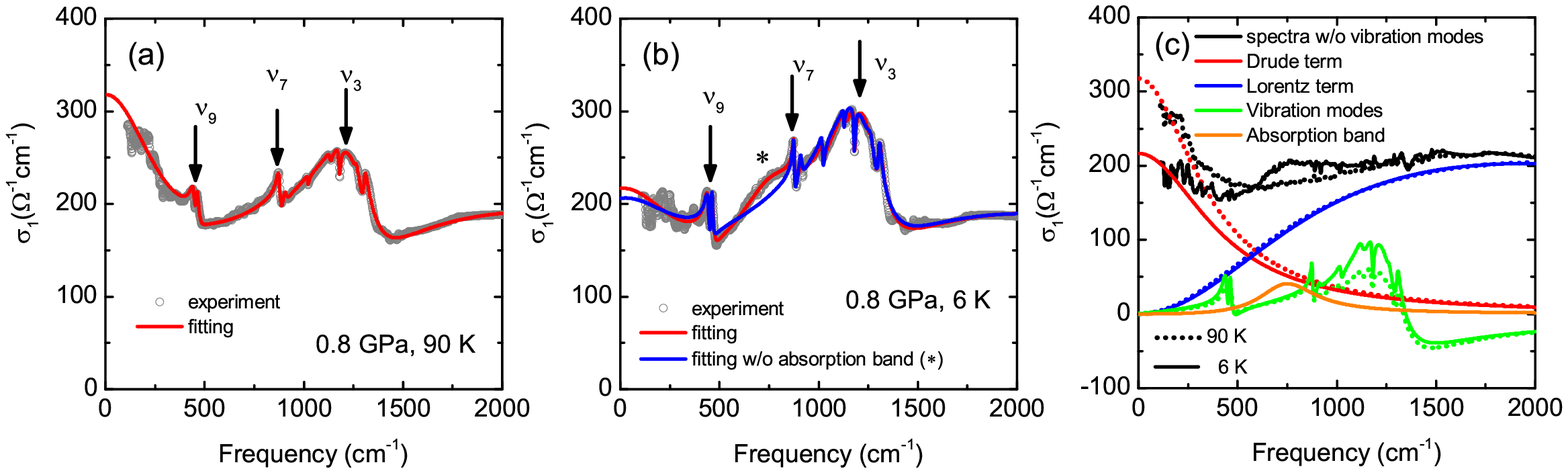}%
\caption{0.8~GPa optical conductivity and fittings at (a) 90~K, above the pseudogap behavior and (b) 6~K, below the pseudogap behavior. Blue fitting in (b) is the best fitting to the experiment without taking into account the broad absorption feature due to pseudogap behavior. (c) shows the vibration modes subtracted spectra as given in the manuscript and the individual contributions for each temperature: Drude (red), high energy Lorentz (blue) and vibration modes (green) for 90~K (dotted) and 6~K (solid) and additional absorption feature (orange) for the 6~K spectrum.}%
\label{fitting}%
\end{figure}

Due to the complex structure of this material many small vibration modes are expected [S4, S5]. We can consistently resolve the ones with high enough contributions. Especially three of these vibration modes are quite big and contribute to the spectra significantly, namely, $\nu_3$, $\nu_7$, and $\nu_9$ modes. In Fig.~\ref{fitting} (a,b), we demonstrated the overall fitting of our 0.8~GPa spectra for two different temperatures above and below the pseudogap behavior. In Fig.\ref{fitting}(c), the spectra witout the vibration modes (as given in the manuscript) is given with the individual contributions for each temperature:  Drude (red), high energy Lorentz (blue) and vibration modes (green) for 90~K (dotted) and 6~K (solid) and additional absorption feature (orange) for the 6~K spectrum. 

The energy range of the determined vibration modes are fit well with the literature. Moreover, they show sharpening with decreasing temperature, as one would expect. In Fig.~\ref{fitting}(b), we also demonstrated the best fitting to the experiment without taking into account the broad absorption feature due to the pseudogap behavior. As one can see, it is not possible to reproduce our spectrum without this broad mode. 

\section{Charge-order boundary}

The ambient and low pressure regime (up to 0.8~GPa) show  a charge ordered (CO) ground state at low temperatures, while the high pressure regime is distinctly different with a clear Drude-like response of the trivial bands and $\omega$-independent $\sigma_1(\omega)$ giving evidence the existance of Dirac electrons. On the other hand, 0.8~GPa spectra can be classified neither like the low pressure regime nor like the high pressure range.

\begin{figure}[h]%
\centering
\includegraphics[scale=1.2]{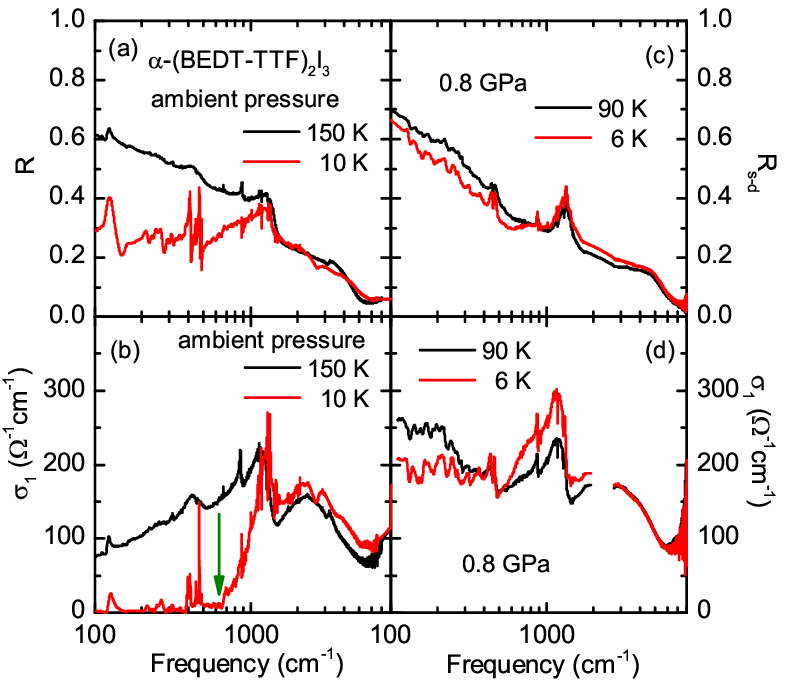}%
\caption{Reflectivity (a,c) and optical conductivity (b,d) of \aeti\ at ambient pressure (left) and 0.8~GPa (right). }%
\label{COboundry}%
\end{figure}

As can be seen from Fig.~\ref{COboundry}, at ambient pressure, at 150~K just above the CO transition, the absolute value of the reflectivity increases steeply with decreasing frequency in the low energy range, a typical semi-metallic behavior. Below the CO temperature ($T_{CO}$) the behavior changes to the insulating one with a sudden drop of the reflectivity at the low energy range and the vibration modes become sharper. The calculated $\sigma_1(\omega)$ can describe this first order transition very well: The non-zero but relatively small SW below 400~\cm\ at 150K is suppressed when the first order transition are reached at $T_{CO}$ with a clear gap opening (marked with green arrow in Fig.~\ref{COboundry}(b)). As for the 0.8 GPa spectra, disappearence of the sharp drop of the optical conductivity suggest that the charge ordering is not the case for this pressure, while the lack of $\omega$-independent optical conductivity indicate that the system is not in the Dirac regime, either. The existence of the significantly large vibration mode at around 1300~\cm\ compared to the higher pressure range further support that this pressure is somewhat very close to the crossover regime.\\

\begin{figure}[h]%
\centering
\includegraphics[scale=1.2]{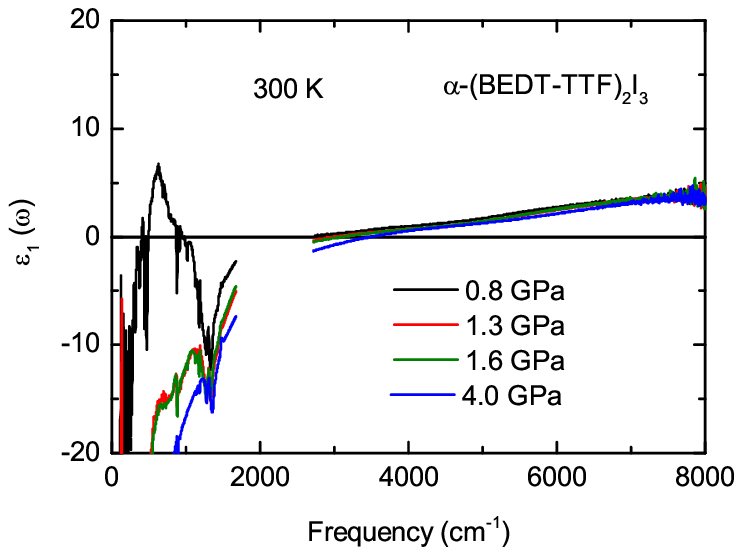}%
\caption{Pressure-dependent real part of the dielectric constant ($\epsilon_1$) of \aeti\ at room temperature.}%
\label{epsinf}%
\end{figure}

\section{Determination of $\epsilon_{\infty}$}

As given in the manuscript, the theoretical excitonic gap estimation requires one to have knowledge of $\epsilon_{\infty}$, which can be estimated from the optical conductivity spectra. $\epsilon_{\infty}$ can be defined as the contribution of the higher energy optical transitions to $\epsilon_1$ (real part of the permittivity), which one can obtain from optical conductivity as $\epsilon_1(\omega) = 1-4\pi\sigma_2(\omega)/\omega$. While one can obtain the real $\sigma_1(\omega)$ and imaginary $\sigma_2(\omega)$ part of the optical conductivity from measured reflectivity via Kramers-Kronig analysis. The obtained $\epsilon_1(\omega)$ is given in Fig.~\ref{epsinf} for various pressures at room temperature. The value of $\epsilon_1(\omega)$ at the high energy limit is $\approx$~4, which can be taken as $\epsilon_{\infty}$. The temperature-independent optical conductivity at high energy range  (Fig.~\ref{RefCon} (e-h)) suggests this value does not change with temperature as well.\\\\

\section*{Supplementary references}
\begin{small}
\setlength\parindent{0pt}[S1] H. K. Mao, J. Xu, and P. M. Bell, J. Geophys. Res. Solid Earth \textbf{91}, 4673 (1986).

\setlength\parindent{0pt}[S2] C. A. Kuntscher, A. Huber, and M. H\"{u}cker, Phys. Rev. B \textbf{89}, 134510 (2014).

\setlength\parindent{0pt}[S3] A. Pashkin, M. Dressel, and C. A. Kuntscher, Phys. Rev. B \textbf{74}, 165118 (2006).

\setlength\parindent{0pt}[S4] Y. Yue, K. Yamamoto, M. Uruichi, C. Nakano, K. Yakushi, S. Yamada, T. Hiejima, and A. Kawamoto, Phys. Rev. B \textbf{82}, 075134

\setlength\parindent{17pt}(2010).

\setlength\parindent{0pt} [S5] T. Ivek, B. Korin-Hamzi\'{c}, O. Milat, S. Tomi\'{c}, C. Clauss, N. Drichko, D. Schweitzer, and M. Dressel, Phys. Rev. B \textbf{83}, 165128 

\setlength\parindent{17pt} (2011).\\

\end{small}

\end{document}